**Fast computation of soft tissue thermal response under deformation based on fast explicit dynamics finite element algorithm for surgical simulation**

Jinao Zhang, Sunita Chauhan

Department of Mechanical and Aerospace Engineering, Monash University, Wellington Road, Clayton, VIC 3800, Australia

**Abstract.**

*Background and Objectives*: During thermal heating surgical procedures such as electrosurgery, thermal ablative treatment and hyperthermia, soft tissue deformation due to surgical tool-tissue interaction and patient movement can affect the distribution of thermal energy induced. Soft tissue temperature must be obtained from the deformed tissue for precise delivery of thermal energy. However, the classical Pennes bio-heat transfer model can handle only the static non-moving state of tissue. In addition, in order to enable a surgeon to visualize the simulated results immediately, the solution procedure must be suitable for real-time thermal applications.

*Methods*: This paper presents a formulation of bio-heat transfer under the effect of soft tissue deformation for fast or near real-time tissue temperature prediction, based on fast explicit dynamics finite element algorithm (FED-FEM) for transient heat transfer. The proposed thermal analysis under deformation is achieved by transformation of the unknown deformed tissue state to the known initial static state via a mapping function. The appropriateness and effectiveness of the proposed formulation are evaluated on a realistic virtual human liver model with blood vessels to demonstrate a clinically relevant scenario of thermal ablation of hepatic cancer.

*Results*: For numerical accuracy, the proposed formulation can achieve a typical $10^{-3}$ level of normalized relative error at nodes and between $10^{-4}$ and $10^{-5}$ level of total errors for the simulation, by comparing solutions against the commercial finite element analysis package. For computation time, the proposed formulation under tissue deformation with anisotropic temperature-dependent properties consumes $2.518 \times 10^{-4}\ ms$ for one element thermal loads computation, compared to $2.237 \times 10^{-4}\ ms$ for the formulation without deformation which is 0.89 times of the former. Comparisons with three other formulations for isotropic and temperature-independent properties are also presented.

*Conclusions*: Compared to conventional methods focusing on numerical accuracy, convergence and stability, the proposed formulation focuses on computational performance for fast tissue thermal analysis. Compared to the classical Pennes model that handles only the static state of tissue, the proposed formulation can achieve fast thermal analysis on deformed states of tissue and can be applied in addition to tissue deformable models for non-linear heating analysis at even large deformation of soft tissue, leading to great translational potential in dynamic tissue temperature analysis and thermal dosimetry computation for computer-integrated medical education and personalized treatment.

The source code is available at https://github.com/jinaojakezhang/FEDFEMBioheatDeform.

## 1. Introduction

Fast computation of thermal response of soft biological tissue is increasingly important in the application of computer-assisted surgery for surgical training, planning and guidance [1]. It can be utilized to (i) enable virtual-reality or augmented-reality-based medical simulators for surgical training, such as training of electrocautery procedures in laparoscopic cholecystectomy [2] and laparoscopic rectal cancer radical surgery [3], (ii) to achieve fast or near real-time surgical planning and optimization of tissue temperature field for hyperthermia treatments [4] and (iii) to improve surgical feedback in thermal ablative cancer treatments [5] by providing temperature indication and thermal damage evaluation [6], leading to technical innovation in medical education and clinical procedures and ultimately improvement of patient outcomes [7].

During thermal heating procedures such as electrosurgery, thermal ablative treatment and hyperthermia, soft tissue can deform due to surgical tool-tissue interaction [8] and patient movement [9] which can affect the distribution of induced thermal energy. The clinical outcome of these surgical procedures relies on accurate control of thermal energy to induce a desired thermal injury to the target tissue without affecting nearby healthy tissue/organs; either excessive or insufficient thermal energy will compromise the effectiveness of the treatment. Therefore, for precise control of induced thermal energy, soft tissue temperature must be obtained from the deformed tissue.

It is challenging to achieve fast computation of thermal response on deformed soft tissue. It was reported that (i) soft tissue thermal properties such as tissue density, thermal conductivity and specific heat capacity vary with temperature [10, 11], and (ii) the effect of bio-heat transfer is due to the combined thermal effect of bio-heat conduction, blood perfusion, metabolic heat generation and externally induced heat [12]; both points lead to non-linear characteristics of bio-heat transfer in soft tissue, resulting in a non-linear system of equations that is computationally expensive to solve via the conventional iterative Newton-Raphson's procedure at each time step [13]. Furthermore, the classical Pennes bio-heat transfer model [12] is based on a static non-moving state of tissue [14], but soft tissue undergone deformation must be accommodated when computing the non-linear tissue temperature field.

Various methods were reported for fast solutions of bio-heat transfer models; however, the study on fast thermal analysis under tissue deformation is very limited. Studies were focused on facilitating the computational efficiency by using parallel alternating direction explicit scheme [15] based on finite difference method (FDM) [16], spatial filter method based on Fast Fourier Transform (FFT) [17], fast FFT method [18], Graphics Processing Unit (GPU)-accelerated FDM [19, 20], GPU-accelerated finite element methodology (FEM) [21], cellular neural network [22], multi-grid technique based on finite volume method (FVM) [23, 24], dynamic mode decomposition based on meshless point collocation method [25] and model order reduction based on FDM [26]. Despite the improved computational effort by the above methods, they all consider solving the bio-heat transfer equation on a static non-moving state of soft tissue. In contrast, Schwenke et al. [14] considered soft tissue deformation and presented a GPU-accelerated FDM for fast simulation of focused ultrasound treatments during respiratory motion; however, FDM requires a regular computation grid occupying the problem domain for spatial derivatives, but human tissue/organs are irregular shapes with curvilinear boundaries, leading to difficulties in accurate computation of tissue material properties and boundary conditions. Furthermore, the utilization of GPU only improves the computational effort of FDM due to parallel solution procedures on hardware but does not solve the computational efficiency fundamentally [27]. To demonstrate the importance of including tissue deformation in dosage computation, Brock et al. [28] compared the average change in prescribed dose between a static model, a model including rigid body motion and a model including rigid motion and deformation. Their study was focused on radiation dose but has shown the model incorporating tissue deformation achieved the least difference to the prescribed dose for the patient's actual tumor. Their study did not consider fast numerical computation.

Recently, the authors studied a fast explicit dynamics finite element algorithm (FED-FEM) [29] for fast transient non-linear heat transfer analysis and extended the algorithm to fast computation of classical Pennes bio-heat transfer problems in the static non-moving tissue state [30]. Compared to classical FEMs focusing on numerical accuracy, convergence and stability, FED-FEM addresses the computational efficiency of thermal analysis to be suitable for fast or near real-time-based engineering applications such as surgical simulation. It employs computationally efficient explicit dynamics in the temporal domain and explicit FEM formulation, element-level computation and low-order finite element formulations in the spatial domain and pre-computation of resulting numerical constants, collectively achieving a very efficient algorithm that is well suited for real-time thermal applications. Compared to classical non-linear FEM, FED-FEM is based on finite element mesh for fast thermal analysis but without classical FEM matrix assembly and iterative computation; there is no need for forming the non-linear thermal stiffness matrix for the entire model in FED-FEM, and a time step is performed directly without the need for the iterative Newton-Raphson's procedure anywhere in the algorithm, together with benefits of pre-computing constant simulation parameters, straightforward treatment of non-linearities and independent equations that are well suited for parallel implementation.

This paper further contributes to the authors' work [30] to enable fast tissue thermal analysis under deformation for surgical simulation. Compared to the classical Pennes model that handles only the static state of tissue, new formulations of FED-FEM are developed to obtain tissue thermal responses on the deformed state of tissue while achieving fast numerical computation compared to classical non-linear FEMs for real-time surgical simulation. It formulates the thermal computation by a transformation of the unknown deformed configuration of tissue to the known initial static non-moving configuration via a mapping function in the form of deformation gradient, which can accommodate large strains at large deformation of soft tissue. The clinical relevance of the presented formulation is demonstrated using a realistic vascularized virtual human liver model for simulation of a thermal ablative treatment of hepatic cancer with soft tissue deformation. Numerical accuracy and computational performance are evaluated.

The remainder of this paper is organized as follows: Section 2 presents the proposed modelling of the Pennes bio-heat transfer model under the effect of soft tissue deformation; Section 3 presents proposed solution procedures based on FED-FEM, and simulated results are evaluated in Section 4; discussions are presented in Section 5, and finally the paper concludes in Section 6 with future improvements of the work.

## 2. Modelling of Pennes bio-heat transfer model under soft tissue deformation

Soft tissue undergoes deformation when experiencing externally applied forces, and tissue temperature must be computed from the deformed tissue. During electrosurgery, hyperthermia and thermal ablative treatment, soft tissue/organs are deformed due to external forces such as the contact between surgical tools and tissue and the patient movements (e.g., respiration and heart motion); the deformation of soft tissue affects the distribution of induced thermal energy, and hence it must be accommodated for tissue temperature prediction for precise control of thermal energy.

Pennes bio-heat transfer (PBHT) model [12] is employed for governing the variation of temperature in soft tissue. The heat transfer in soft biological tissue can be characterized by various bio-heat transfer models, such as PBHT [12], Wulff model [31], Klinger model [32], Chen and Holmes model [33] and discrete vasculature model [34] (see review paper [35]), among which PBHT [12] is widely applied [36-39] and has been validated against experimental data [40-42] to provide reliable tissue temperature predictions. The governing equation of three-dimensional (3-D) non-linear transient PBHT accounting for bio-heating and cooling effects of anisotropic tissue heat conduction, isotropic blood perfusion, metabolic heat generation and regional heat sources is given by

$$ {}^{t}\rho\ {}^{t}c \frac{\partial\ {}^{t}T(\mathbf{x})}{\partial t} = \nabla \cdot \left( {}^{t}k \nabla\ {}^{t}T(\mathbf{x}) \right) - {}^{t}w_b\ {}^{t}c_b \left( {}^{t}T(\mathbf{x}) - {}^{t}T_a \right) + {}^{t}Q + {}^{t}H \quad \forall \mathbf{x} \in \Omega \tag{1} $$

where ${}^{t}\rho$ is the tissue mass density $[kg/m^3]$, ${}^{t}c$ the tissue specific heat capacity $[J/(kg \cdot K)]$, ${}^{t}T(\mathbf{x})$ the local tissue temperature $[K]$ at 3-D spatial point $\mathbf{x}(x, y, z)$ at time $t$ $[s]$, ${}^{t}k$ the tissue thermal conductivity $[W/(m \cdot K)]$, ${}^{t}w_b$ the volumetric blood perfusion rate $[kg/(m^3 \cdot s)]$, ${}^{t}c_b$ the blood specific heat capacity $[J/(kg \cdot K)]$, ${}^{t}T_a$ the arterial blood temperature $[K]$, ${}^{t}Q$ the metabolic heat generation rate $[W/m^3]$, ${}^{t}H$ the heat rate of external applied heat source $[W/m^3]$ that represents the heat applied during hyperthermia or thermal ablative treatments, $\nabla \cdot$ the divergence operator and $\nabla$ the gradient operator on the 3-D spatial domain $\Omega\left(\mathbf{x}(x, y, z)\right) \subset \mathbb{R}^3$ ; the state variables are, in general, temperature-dependent [10, 11, 43] and can be directly accounted for in the above equation for transient tissue thermal analysis; for indication of state variables in the relevant system configuration: (i) a left superscript denotes the system configuration in which a quantity occurs, and (ii) a left subscript (if applicable) denotes the configuration with respect to which the quantity is measured (0 denotes the initial system configuration and $t$ the current system configuration at time $t$), this notation will be used throughout the present work.

The classical PBHT was calculated for tissue temperature only in the static non-moving state of tissue [14]. In order to apply PBHT on the deformed soft tissue, the proposed methodology employs a transformation of the unknown deformed system configuration to the known initial static non-moving system configuration via a mapping function. Using the deformed system configuration ${}^{t}\Omega$ at time $t$, PBHT can be integrated over tissue volume to get

$$ \int_{{}^{t}\Omega} \left( {}^{t}\rho\ {}^{t}c \frac{\partial\ {}^{t}T(\mathbf{x})}{\partial t} \right) d\ {}^{t}\Omega = \int_{{}^{t}\Omega} \left( {}^{t}\nabla \cdot \left( {}^{t}k\ {}^{t}\nabla\ {}^{t}T(\mathbf{x}) \right) - {}^{t}w_b\ {}^{t}c_b \left( {}^{t}T(\mathbf{x}) - {}^{t}T_a \right) + {}^{t}Q + {}^{t}H \right) d\ {}^{t}\Omega \tag{2} $$

Considering a transformation of system configuration ${}^{0}\Omega \rightarrow {}^{t}\Omega$, the mapping function can be defined by

$$ \xi_t :\ {}^{0}\Omega \times \mathbb{R}_0^+ \rightarrow {}^{t}\Omega, \qquad {}^{t}\mathbf{x} = \xi_t\left( {}^{0}\mathbf{x} \right) = \xi\left( {}^{0}\mathbf{x}, t \right) \tag{3} $$

where ${}^{0}\mathbf{x}$ and ${}^{t}\mathbf{x}$ are the spatial coordinates of material point $\mathbf{x}(x, y, z) \in \Omega$ in the initial ${}^{0}\Omega$ and current ${}^{t}\Omega$ system configurations at time 0 and $t$, respectively; the mapping function $\xi$ is continuous in time, orientation-preserving, continuously differentiable and globally invertible at all times.

By performing a change of variables, the volume-integral can be expressed in the form of the known initial system configuration ${}^{0}\Omega$ as

$$ \int_{{}^{t}\Omega} (\ldots)\, d\ {}^{t}\Omega = \int_{{}^{0}\Omega} (\ldots) \det\left( \mathcal{L}\xi\left( {}^{0}\mathbf{x}, t \right) \right) d\ {}^{0}\Omega = \int_{{}^{0}\Omega} (\ldots) \det\left( \frac{\partial\ {}^{t}\mathbf{x}}{\partial\ {}^{0}\mathbf{x}} \right) d\ {}^{0}\Omega \tag{4} $$

where $\mathcal{L}$ is the differential operator.

By applying the chain rule, the heat conduction term in the classical PBHT can be expressed in the form of the initial system configuration ${}^{0}\Omega$ as

$$\int_{{}^{t}\Omega} {}^{t}\nabla \cdot \left( {}^{t}k \; {}^{t}\nabla \; {}^{t}T(\mathbf{x}) \right) d \; {}^{t}\Omega = \int_{{}^{0}\Omega} \left( {}^{0}\nabla \left( \frac{\partial \; {}^{0}\mathbf{x}}{\partial \; {}^{t}\mathbf{x}} \right) \cdot \left( {}^{t}k \; {}^{0}\nabla \left( \frac{\partial \; {}^{0}\mathbf{x}}{\partial \; {}^{t}\mathbf{x}} \right) \; {}^{t}T(\mathbf{x}) \right) \right) \det \left( \frac{\partial \; {}^{t}\mathbf{x}}{\partial \; {}^{0}\mathbf{x}} \right) d \; {}^{0}\Omega$$

$$= \int_{{}^{0}\Omega} \left( {}^{0}\nabla {}_{0}^{t}\mathbf{F}^{-1} \cdot \left( {}^{t}k \; {}^{0}\nabla {}_{0}^{t}\mathbf{F}^{-1} \; {}^{t}T(\mathbf{x}) \right) \right) \det({}_{0}^{t}\mathbf{F}) \, d \; {}^{0}\Omega \tag{5}$$

where ${}_{0}^{t}\mathbf{F}$ is the deformation gradient tensor which is the fundamental measure of deformation and is given by

$${}_{0}^{t}\mathbf{F} = \frac{\partial \; {}^{t}\mathbf{x}}{\partial \; {}^{0}\mathbf{x}} \tag{6}$$

where the displacement from ${}^{0}\mathbf{x}$ to ${}^{t}\mathbf{x}$ from time 0 to $t$ is governed by the non-rigid mechanics of motion for soft tissue deformation, i.e.,

$${}^{t}\rho \frac{\partial^2 \; {}^{t}\mathbf{x}}{\partial t^2} + {}^{t}\gamma \frac{\partial \; {}^{t}\mathbf{x}}{\partial t} + \frac{\delta \; {}^{t}W(\mathbf{x})}{\delta \; {}^{t}\mathbf{x}} = {}^{t}\mathbf{f}(\mathbf{x}) \tag{7}$$

where ${}^{t}\gamma$ is the system damping constant, ${}^{t}W(\mathbf{x})$ the instantaneous strain energy and ${}^{t}\mathbf{f}(\mathbf{x})$ the net external force at $\mathbf{x}$.

It is worth noting that Eq. (5) is based only on three state variables at time $t$: (i) the deformation gradient ${}_{0}^{t}\mathbf{F}$ from initial system configuration ${}^{0}\Omega$ to deformed configuration ${}^{t}\Omega$, (ii) the thermal conductivity ${}^{t}k$ which may be temperature-in/dependent and (iii) the tissue temperature ${}^{t}T(\mathbf{x})$ at the deformed configuration ${}^{t}\Omega$; the other state variables are defined over the known static non-moving initial configuration ${}^{0}\Omega$, and hence they can be pre-computed for efficient run-time computation. Numerically, the above formulation of PBHT accommodates large strains with non-linear material and geometric properties of bio-heat transfer at large deformation of soft tissue. It is also compatible with small/infinitesimal deformation of soft tissue, which can be described by the linear deformation theory in which the variation of deformation gradient to unity is not significant and may be neglected.

## 3. Numerical solution procedures

FED-FEM is introduced in Section 3.1. The proposed fast thermal solution procedure with incorporation of tissue deformation based on FED-FEM is presented in Section 3.2.

### 3.1 Fast explicit dynamics finite element algorithm (FED-FEM)

The proposed solution procedure is based on FED-FEM [29], which is suitable for real-time thermal applications and has been applied to fast/real-time analysis of static non-moving transient PBHT problems [30]. The main attributes of FED-FEM for fast computation are:

- explicit time integration,
- nodal loads computation at element-level,
- computationally efficient element formulation,
- explicit formulation for unknown state variables, and
- pre-computation of resulting simulation constants.

Collectively, the above attributes lead to significant computation time reductions in both temporal and spatial domains owning to:

- eliminating the necessity for system stiffness inversion at each time step→fast computation of time derivative of state variables ($\frac{\partial \; {}^{t}T(\mathbf{x})}{\partial t}$),
- eliminating the necessity of assembling non-linear system stiffness matrices for the entire model→fast computation of nodal loads (thermal loads),
- computing non-linear system responses efficiently in spatial domain→fast computation of nodal loads (thermal loads),
- eliminating the necessity for iterations and allowing straightforward parallel computation→fast computation of nodal state variables ( ${}^{t}T(\mathbf{x})$), and

- improving computational efficiency at run-time simulation → fast computation of nodal state variables ( ${}^{t}T(\mathbf{x})$).

Applying FED-FEM in the known static non-moving initial system configuration further enables the following computational benefits:

- pre-computation of all known state variables, such as spatial derivatives, at the reference configuration, and
- numerical errors due to consecutive spatial interpolations do not accumulate from one time step to another.

Compared to classical non-linear FEM where solutions are sought by assembling and inverting global thermal matrices (e.g., non-linear thermal stiffness matrix) for temperature at future time points, FED-FEM does not require assemblies of non-linear thermal matrices for the entire model; instead, it directly computes nodal non-linear thermal loads based on non-linear element thermal stiffness; the treatment of boundary conditions is performed at the element-level computation in the form of nodal heat flows, instead of assembling into the global thermal matrices in classical FEM. Furthermore, as a result of the collective attributes of FED-FEM, it leads to constant simulation matrices (e.g., initial element Jacobian and temperature gradient matrix) that can be pre-computed. Advancing in time requires only one computation per node without iterative Newton-Raphson's procedure. These key attributes make FED-FEM well suited for fast/real-time solutions of transient response of the proposed model.

## 3.2 Proposed FED-FEM formulation for PBHT under soft tissue deformation

Using the first-order explicit forward time integration, the spatially discretized matrix equation of PBHT can be written as

$$ {}^{t}\mathbf{C}\left(\frac{{}^{t+\Delta t}\mathbf{T}(\mathbf{x}) - {}^{t}\mathbf{T}(\mathbf{x})}{\Delta t}\right) = {}_{0}^{t}\mathbf{K}\ {}^{t}\mathbf{T}(\mathbf{x}) - {}^{t}\mathbf{K}_{b}\ {}^{t}\mathbf{T}(\mathbf{x}) + {}^{t}\mathbf{G}_{b} + {}^{t}\mathbf{Q} + {}^{t}\mathbf{H} \tag{8} $$

where ${}^{t}\mathbf{C}$ is the thermal mass (mass and specific heat capacity) matrix, ${}^{t+\Delta t}\mathbf{T}(\mathbf{x})$ the vector of nodal temperatures at time $t + \Delta t$ with $\Delta t$ the time step, ${}_{0}^{t}\mathbf{K}$ the thermal stiffness (heat conduction) matrix, ${}^{t}\mathbf{K}_{b}$ the thermal stiffness (blood perfusion) matrix, ${}^{t}\mathbf{G}_{b}$ the vector of heat flows of blood perfusion at arterial blood temperature, ${}^{t}\mathbf{Q}$ the vector of heat flows of metabolic heat generation and ${}^{t}\mathbf{H}$ the vector of heat flows of external heat sources; by employing explicit FEM (lumped mass approximation) while ensuring mass conservation, it leads to a block diagonalized thermal mass matrix ${}^{t}\mathbf{C}^{diag}$ and a block diagonalized thermal stiffness (blood perfusion) matrix ${}^{t}\mathbf{K}_{b}^{diag}$.

The above equation may be further arranged into

$$ {}^{t+\Delta t}\mathbf{T}(\mathbf{x}) = {}^{t}\mathbf{T}(\mathbf{x}) + \Delta t\ {}^{t}\mathbf{C}^{diag^{-1}}\left(\sum_{e} {}_{0}^{t}\hat{\mathbf{F}}_{e} - {}^{t}\mathbf{K}_{b}^{diag}\ {}^{t}\mathbf{T}(\mathbf{x}) + {}^{t}\mathbf{G}_{b} + {}^{t}\mathbf{Q} + {}^{t}\mathbf{H}\right) \tag{9} $$

where

$$ \sum_{e} {}_{0}^{t}\hat{\mathbf{F}}_{e} = {}_{0}^{t}\mathbf{K}\ {}^{t}\mathbf{T}(\mathbf{x}) = {}_{0}^{t}\hat{\mathbf{F}} \tag{10} $$

where ${}_{0}^{t}\hat{\mathbf{F}}_{e}$ are the thermal load components due to bio-heat conduction in element $e$ of the global nodal thermal loads ${}_{0}^{t}\hat{\mathbf{F}}$. For a given element $e$, ${}_{0}^{t}\hat{\mathbf{F}}_{e}$ is computed by

$$ {}_{0}^{t}\hat{\mathbf{F}}_{e} = \int_{{}^{t}V_{e}} \left(\frac{\partial\ {}^{t}\mathbf{h}({}^{t}\mathbf{x})}{\partial\ {}^{t}\mathbf{x}}\right)^{T} {}^{t}\mathbf{D}\left(\frac{\partial\ {}^{t}\mathbf{h}({}^{t}\mathbf{x})}{\partial\ {}^{t}\mathbf{x}}\right) d\ {}^{t}V\ \ {}^{t}\mathbf{T}(\mathbf{x}) \tag{11} $$

where ${}^{t}V_{e}$ is the volume of the $e$th element at time $t$, ${}^{t}\mathbf{h}({}^{t}\mathbf{x})$ the shape functions of the element and ${}^{t}\mathbf{D}$ the element thermal conductivity matrix which may be temperature-in/dependent.

If selecting the shape functions in the initial reference configuration ${}^{0}\Omega$ as ${}^{0}\mathbf{h}({}^{0}\mathbf{x}) = {}^{t}\mathbf{h}({}^{t}\mathbf{x})$ and incorporating it into the proposed formulation, the above equation can be written in terms of the initial system configuration ${}^{0}\Omega$ as

$$
\begin{aligned}
{}^t_0\hat{\mathbf{F}}_e &= \int_{{}^0V_e} \left(\frac{\partial\, {}^t\mathbf{h}({}^t\mathbf{x})}{\partial\, {}^0\mathbf{x}} \frac{\partial\, {}^0\mathbf{x}}{\partial\, {}^t\mathbf{x}}\right)^T {}^t\mathbf{D} \left(\frac{\partial\, {}^t\mathbf{h}({}^t\mathbf{x})}{\partial\, {}^0\mathbf{x}} \frac{\partial\, {}^0\mathbf{x}}{\partial\, {}^t\mathbf{x}}\right) \det\left(\frac{\partial\, {}^t\mathbf{x}}{\partial\, {}^0\mathbf{x}}\right) d\,{}^0V \;\, {}^t\mathbf{T}(\mathbf{x}) \\
&= \int_{{}^0V_e} \left(\frac{\partial\, {}^0\mathbf{h}({}^0\mathbf{x})}{\partial\, {}^0\mathbf{x}}\, {}^t_0\mathbf{F}^{-1}\right)^T {}^t\mathbf{D} \left(\frac{\partial\, {}^0\mathbf{h}({}^0\mathbf{x})}{\partial\, {}^0\mathbf{x}}\, {}^t_0\mathbf{F}^{-1}\right) \det({}^t_0\mathbf{F})\, d\,{}^0V \;\, {}^t\mathbf{T}(\mathbf{x}) \qquad (12) \\
&= \int_{{}^0V_e} \left(\mathcal{L}\, {}^0\mathbf{h}_{\mathbf{x}}\, {}^t_0\mathbf{F}^{-1}\right)^T {}^t\mathbf{D} \left(\mathcal{L}\, {}^0\mathbf{h}_{\mathbf{x}}\, {}^t_0\mathbf{F}^{-1}\right) \det({}^t_0\mathbf{F})\, d\,{}^0V \;\, {}^t\mathbf{T}(\mathbf{x})
\end{aligned}
$$

where ${}^0V_e$ is the initial volume of the $e$th element and $\mathcal{L}\, {}^0\mathbf{h}_{\mathbf{x}}$ the matrix of element shape function spatial derivatives calculated with respect to the known initial reference configuration $\Omega_0$, both of which can be pre-computed to improve computational efficiency at run-time computation. Equations. (9-12) further imply that nodal thermal loads are calculated at the element level, eliminating the necessity for assembling the global non-linear thermal stiffness matrix ${}^t_0\mathbf{K}$ and multiplying temperature vector ${}^t\mathbf{T}(\mathbf{x})$ for the entire model; hence, the computational cost is significantly lower than standard FEM.

The deformation gradient ${}^t_0\mathbf{F}$ for an element can be written in terms of element nodal displacements matrix ${}^t\mathbf{u}_e$ as

$$
{}^t_0\mathbf{F} = \left({}^t\mathbf{u}_e\right)^T \mathcal{L}\, {}^0\mathbf{h}_{\mathbf{x}} + \mathbf{I} \qquad (13)
$$

To enable fast computation of nodal thermal loads, the computationally efficient 3-D eight-node reduced integration hexahedral element and four-node linear tetrahedral element are used. The new formulation of element nodal thermal loads for the eight-node reduced integration hexahedral element is:

$$
{}^t_0\hat{\mathbf{F}}_e = 8\det\left({}^0\mathbf{J}\right)\left(\mathcal{L}\, {}^0\mathbf{h}_{\mathbf{x}}\, {}^t_0\mathbf{F}^{-1}\right)^T {}^t\mathbf{D}\left(\mathcal{L}\, {}^0\mathbf{h}_{\mathbf{x}}\, {}^t_0\mathbf{F}^{-1}\right)\det({}^t_0\mathbf{F})\; {}^t\mathbf{T}(\mathbf{x}) \qquad (14)
$$

where ${}^0\mathbf{J}$ the element Jacobian matrix which can be pre-computed (see Ref. [44] for details).

The new formulation for the four-node linear tetrahedral element is:

$$
{}^t_0\hat{\mathbf{F}}_e = {}^0V_{tet}\left(\mathcal{L}\, {}^0\mathbf{h}_{\mathbf{x}}\, {}^t_0\mathbf{F}^{-1}\right)^T {}^t\mathbf{D}\left(\mathcal{L}\, {}^0\mathbf{h}_{\mathbf{x}}\, {}^t_0\mathbf{F}^{-1}\right)\det({}^t_0\mathbf{F})\; {}^t\mathbf{T}(\mathbf{x}) \qquad (15)
$$

where ${}^0V_{tet}$ is the initial volume of a tetrahedral element which can be pre-computed; at each time step the nodal thermal loads ${}^t_0\hat{\mathbf{F}}_e$ are updated by considering only the variations of deformation gradient ${}^t_0\mathbf{F}$, temperature-dependent thermal conductivity ${}^t\mathbf{D}$ and nodal temperatures ${}^t\mathbf{T}(\mathbf{x})$; other variables can be pre-computed. The deformation of soft tissue, as described by the deformation gradient, is therefore directly incorporated into the element-level computation for nodal thermal loads for temperature prediction.

At each time step, the temperature ${}^{t+\Delta t}T_{(i)}(\mathbf{x})$ at node $i$ can be obtained by

$$
{}^{t+\Delta t}T_{(i)}(\mathbf{x}) = {}^tT_{(i)}(\mathbf{x}) + \Delta t\; {}^tC_{(i)}^{diag^{-1}} \left({}^t_0\hat{F}_{(i)} - {}^tK_{b(i)}^{diag}\; {}^tT_{(i)}(\mathbf{x}) + {}^tG_{b(i)} + {}^tQ_{(i)} + {}^tH_{(i)}\right) \qquad (16)
$$

The temperature-dependent thermal properties and non-linear thermal boundary conditions can be directly accounted for in the above formulation. It is worth noting that the above equation states an explicit formulation to advance unknown temperatures in the temporal domain, leading to a system of uncoupled equations that is particularly suitable for parallel execution. The algorithm of PBHT based on FED-FEM for static non-moving state of soft tissue is presented in Ref. [30]. The modified algorithm in the present work to accommodate soft tissue deformation requires only the nodal displacements for computation of deformation gradient ${}^t_0\mathbf{F}$ (see Eq. (13)), which are readily available from soft tissue deformable algorithms [45].

## 4. Results

The proposed formulation for PBHT under tissue deformation is implemented to evaluate numerical accuracy and computational performance. The clinically relevant application is demonstrated using a realistic vascularized virtual human liver model for simulation of a thermal ablative treatment of hepatic cancer under soft tissue deformation.

### 4.1 Algorithm verification

The proposed methodology is compared against the commercial finite element analysis package, ABAQUS/CAE 2018 (2017_11_08-04.21.41 127140), to verify numerical accuracy on a vascularized human liver model. Owing to irregular geometric shapes of the vascularized liver model with temperature-dependent material properties and complex loading and boundary conditions, the analytical solution does not exist for exact solutions for numerical comparison. Alternatively, ABAQUS/CAE provides an established non-linear finite element analysis procedure and has been applied for numerical verification in our previous works on transient heat transfer [29] and static non-moving PBHT [30]. To evaluate the numerical accuracy of the proposed FED-FEM for PBHT under tissue deformation, the same simulation settings are used for the proposed algorithm and ABAQUS (Dynamic, Temp-disp, Explicit) for the same governing equations for heat transfer and deformable analysis. The numerical convergence of FED-FEM has been previously verified via the standard patch tests in Ref. [29].

#### 4.1.1 Simulation settings

Fig. 1 illustrates the virtual model of the human liver with blood vessel networks (portal veins and vena cava). The 3-D models of the human liver and blood vessels were segmented and constructed from two-dimensional (2-D) Computed Tomography slices. It is worth noting that the classical PBHT is accurate in the absence of large blood vessels [46, 47]. The effect of convective cooling of blood flow in large blood vessels (diameters larger than 0.5 $mm$) is not accounted for in PBHT [48]. To obtain reliable predictions of tissue temperature, the spatial domain $\Omega\big(\mathbf{x}(x,y,z)\big) \subset \mathbb{R}^3$ of the vascularized liver is decomposed into two subdomains: (i) a perfused solid tissue domain $\Omega_s\big(\mathbf{x}(x,y,z)\big) \subset \mathbb{R}^3$ in which the amount of dissipated heat is estimated by volumetric averaging over all hepatic tissues the effect of isotropic and homogeneous blood perfusion due to vascular perfusion on the fine scale of micro capillaries in the liver sinusoids, and (ii) an embedded flowing blood domain $\Omega_b\big(\mathbf{x}(x,y,z)\big) \subset \mathbb{R}^3$ in which the blood is assumed to be not heated considerably due to large volume flows, so that the vessel wall temperature remains the same as the arterial or body core temperature [14].

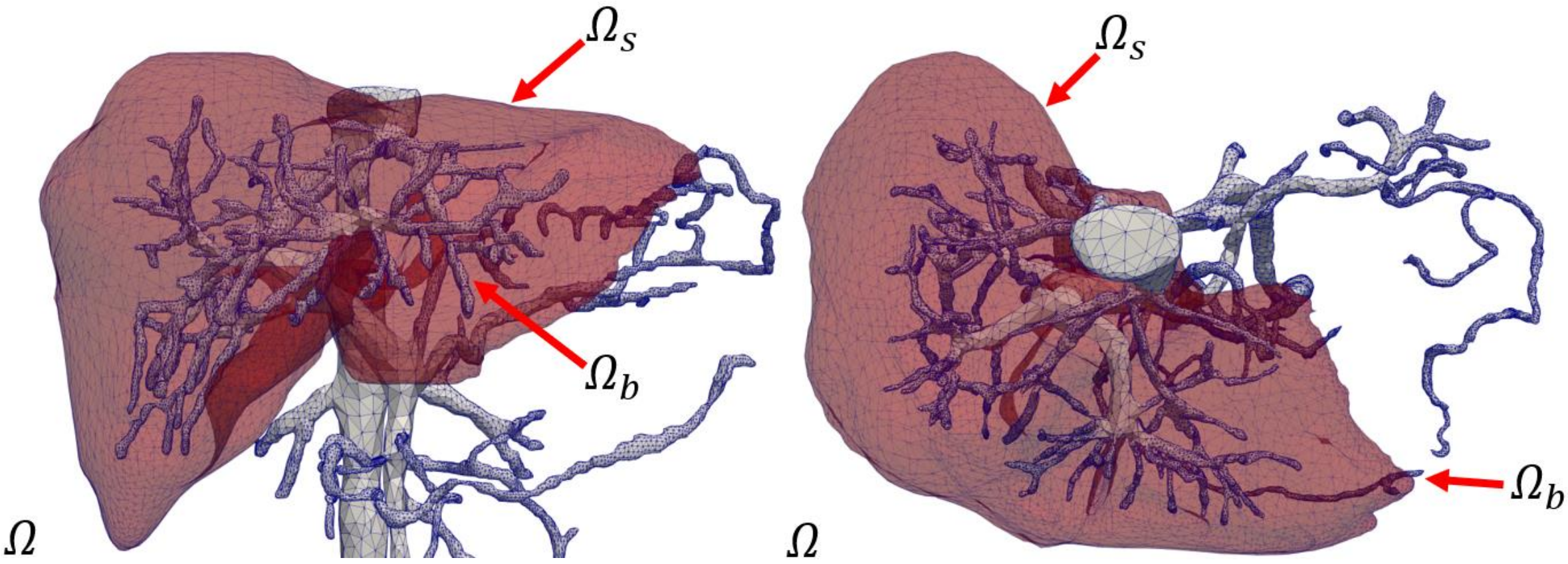

Fig. 1. Virtual models of the human liver with blood vessel networks: the spatial domain $\Omega$ of the vascularized liver is decomposed into a perfused solid tissue domain $\Omega_s$ and an embedded flowing blood domain $\Omega_b$.

The bio-heat transfer in the perfused hepatic solid tissue region $\Omega_s$ is modelled by the proposed PBHT. The bio-heat transfer in the flowing blood region $\Omega_b$ is modelled by enforcing a Dirichlet boundary condition (see Eq. (17)) at the interface of the solid tissue $\Omega_s$ and flowing blood $\Omega_b$ domains with the arterial temperature $T_a$ for the vessel walls. This approach can provide suitable tissue temperature predictions based on studies in Refs. [14, 49]. Thermal ablation performed in the vicinity of the flowing blood region $\Omega_b$ is susceptible to the heat sink effect [50].

$$ {}^{t}T(\mathbf{x}) = {}^{t}T_a \quad \forall \mathbf{x} \in \Omega_b \tag{17} $$

Fig. 2 presents regions (A-E) of the liver model where loading and boundary conditions are prescribed; it is assumed that the liver is subject to induced displacements at region A, the nodes that are in contact with the main large vessel wall (region B) are assumed fixed in position, heat sources are induced in a spherical region denoted by C which can lead to concentrated thermal gradients [29], the arterial temperature is enforced to the flowing blood region D and metabolic heat generation and blood perfusion are prescribed to region E. The simulation settings of the tested liver model, including geometry, loading, boundary and initial conditions, thermal and mechanical material properties and simulation procedures are presented in Table. 1. The employed value for metabolic heat generation is determined based on the metabolic heat generation rate $Q = 33800\ W/m^3$ [51] and liver volume $V = 0.00172\ m^3$ for the 46034 nodes. The employed value for blood perfusion is determined based on the blood perfusion rate $w_b = 26.6\ kg/(m^3 \cdot s)$, blood specific heat $c_b = 3617\ J/(kg \cdot K)$ [37] and liver volume $V = 0.00172\ m^3$ for the 46034 nodes. The associated nodal area is 1 $m^2$ for the concentrated film condition in ABAQUS for blood perfusion. The initial temperature $^0T(\mathbf{x})$ at all nodes and arterial temperature $T_a$ are set to 37 °C. Linear interpolation between points and extrapolation is employed to determine temperature-dependent tissue properties at different temperatures. The liver deformation is simulated using an isotropic neo-Hookean hyperelastic model to approximate the mechanical properties of the human liver. The parameters C10 and D1 for ABAQUS to define the neo-Hookean model were calculated from the Young's modulus and Poisson's ratio. Nodal displacements were recorded and used for computation of deformation gradients (see Eq. (13)) for the proposed algorithm to perform thermal analysis under deformation.

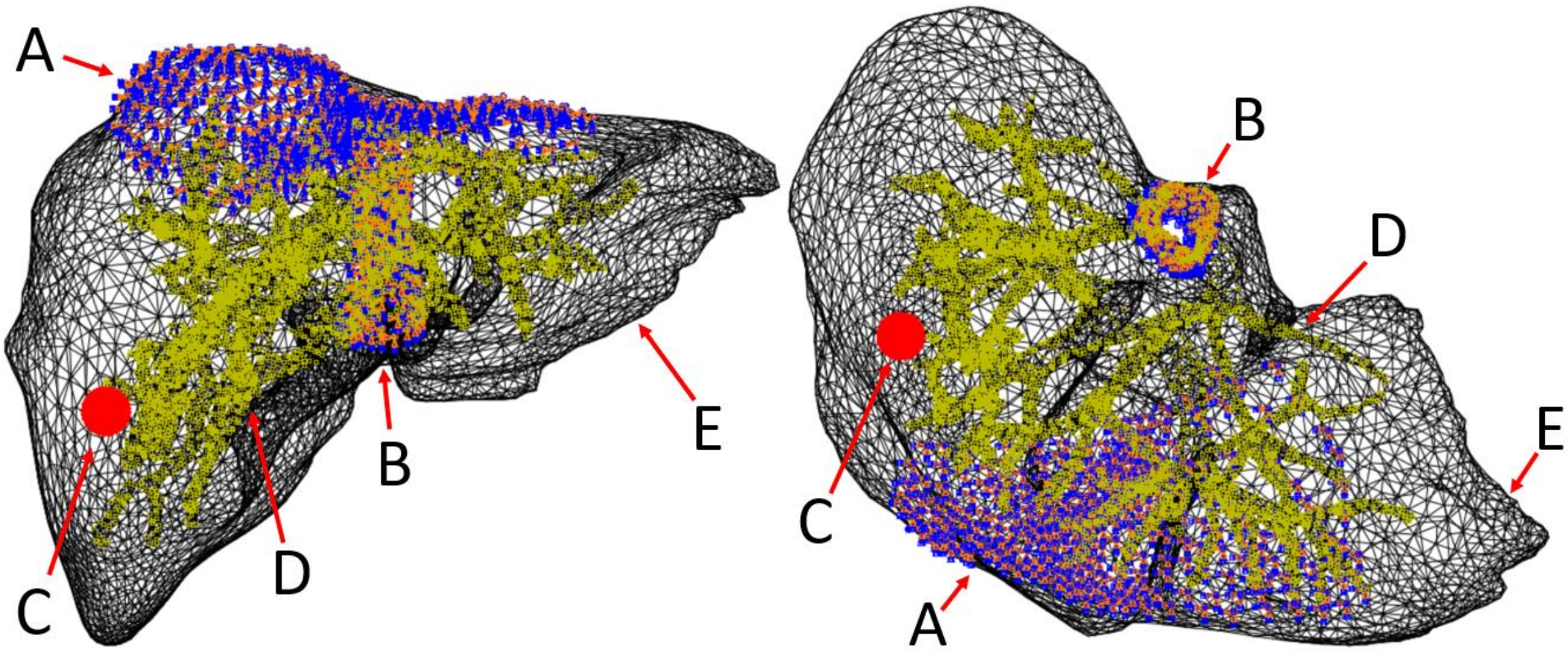

Fig. 2. Regions (A-E) where loading and boundary conditions are prescribed for the simulation, and parameter values are presented in Table. 1.

Table. 1. Simulation settings of the tested liver model.

| Geometry | | Values | Descriptions |
|---|---|---|---|
| Nodes | | 46034 | |
| Degrees of freedom | | 46034 | Thermal: one per node $(T)$ |
| | | 138102 | Mechanical: three per node $(x, y, z)$ |
| Elements | | 257103 | Four-node linear tetrahedrons |
| **Region** | **Conditions** | **Values** | **Descriptions** |
| A | Prescribed displacement | Z: -0.01 $m$ | To induce deformation |
| B | Fixed position | XYZ: 0 $m$ | Nodes on the large blood vessel wall assumed fixed |
| C | Concentrated heat flux | 0.2 $W$/node | Heat sources |
| D | Fixed temperature | 37 °C | Flowing blood region $\Omega_b$ |
| E | Concentrated heat flux | 0.001263 $W$/node | Metabolic heat generation |
| E | Concentrated film | 0.003595 $W/(m^2 \cdot K)$ | Blood perfusion |
| **Material properties** | | **Values** | **Unit** |
| Density | | 1060 | $[kg/m^3]$ |
| Thermal properties | | | |
| • Specific heat capacity | | 3600 @ 37 °C [11] | $[J/(kg \cdot K)]$ |
| | | 3800 @ 65 °C | $[J/(kg \cdot K)]$ |
| • Thermal conductivity | | 0.53 @ 37 °C [11] | $[W/(m \cdot K)]$ |

| | | |
|---|---|---|
| | 0.57 @ 65 ℃ | $[W/(m \cdot K)]$ |
| Mechanical properties (neo-Hookean hyperelastic model) | | |
| • Young's modulus | 3500 [52] | $[Pa]$ |
| • Poisson's ratio | 0.47 | |

| Simulation procedures | Step 1 | Step 2 |
|---|---|---|
| Thermal | 5 s heat source on | 15 s heat source off |
| Mechanical | 10 s displacement on | 10 s displacement off |

### 4.1.2 Results evaluation

The temperature distributions computed by the proposed methodology are illustrated in Fig. 3, and temperature comparisons with ABAQUS for all nodes at time points $t = 5\ s$, $10\ s$, $15\ s$ and $20\ s$ are presented in Fig. 4. It can be seen there is good agreement between those from the proposed method and ABAQUS solutions with the same simulation settings. Fig. 5 further presents temperature differences ($T^{ABAQUS} - T^{Proposed}$) and histograms of the normalized relative errors. There is only a marginal difference between the results obtained from the proposed method and ABAQUS. Fig. 6 presents numerical comparisons of temperature-history between the results of the proposed method (P,N#) and ABAQUS (A,N#) at selected nodes. It is worth noting that the normalized relative error in Fig. 5 is calculated based on the following equation:

$$Normalized\ relative\ error\ = \left| \frac{T_i^{Proposed} - T_i^{Abaqus}}{T_{max}^{Abaqus} - T_{min}^{Abaqus}} \right| \tag{18}$$

and the total error is calculated based on the following equation:

$$Total\ error = \sqrt{\frac{\sum_{i=1}^{n} \left(T_i^{Proposed} - T_i^{Abaqus}\right)^2}{\sum_{i=1}^{n} \left(T_i^{Abaqus}\right)^2}} \tag{19}$$

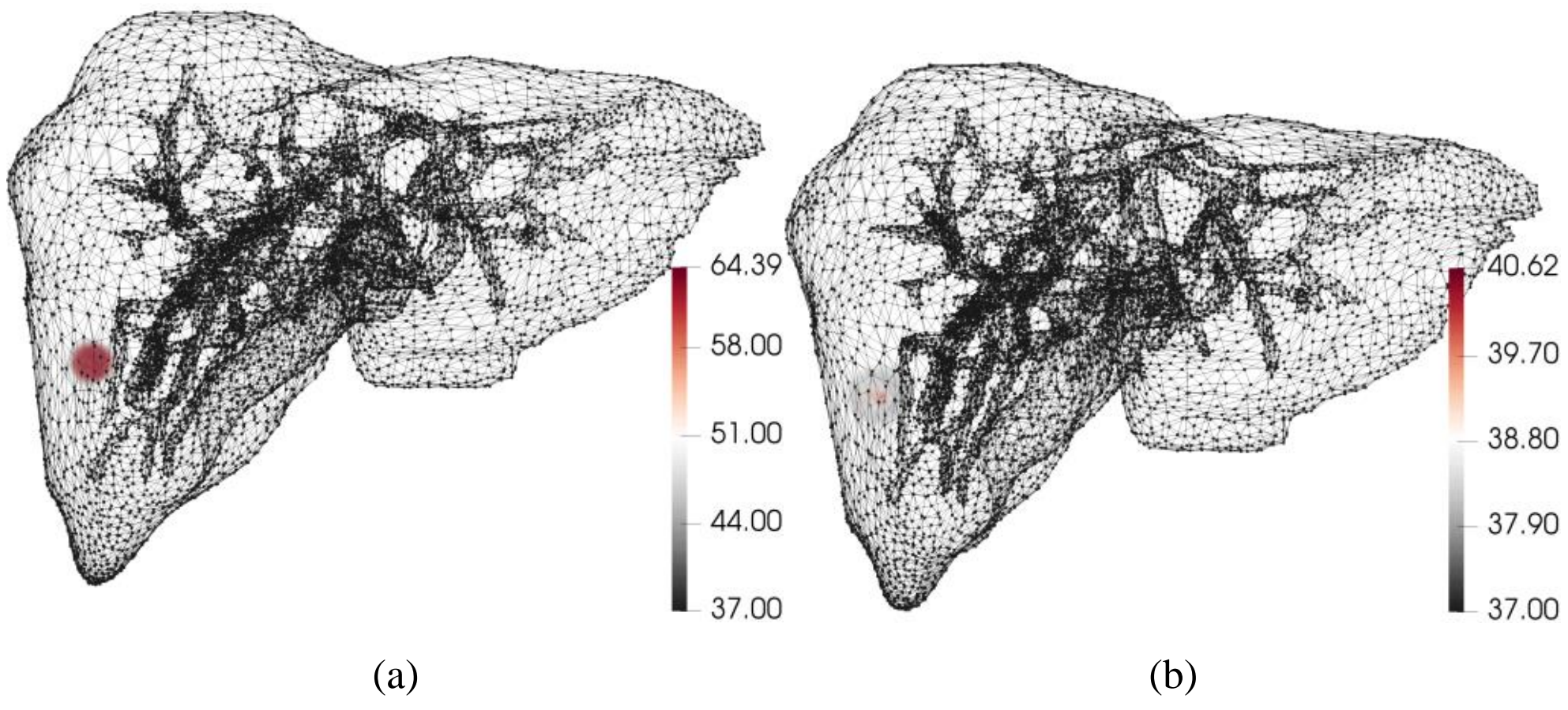

(a) (b)

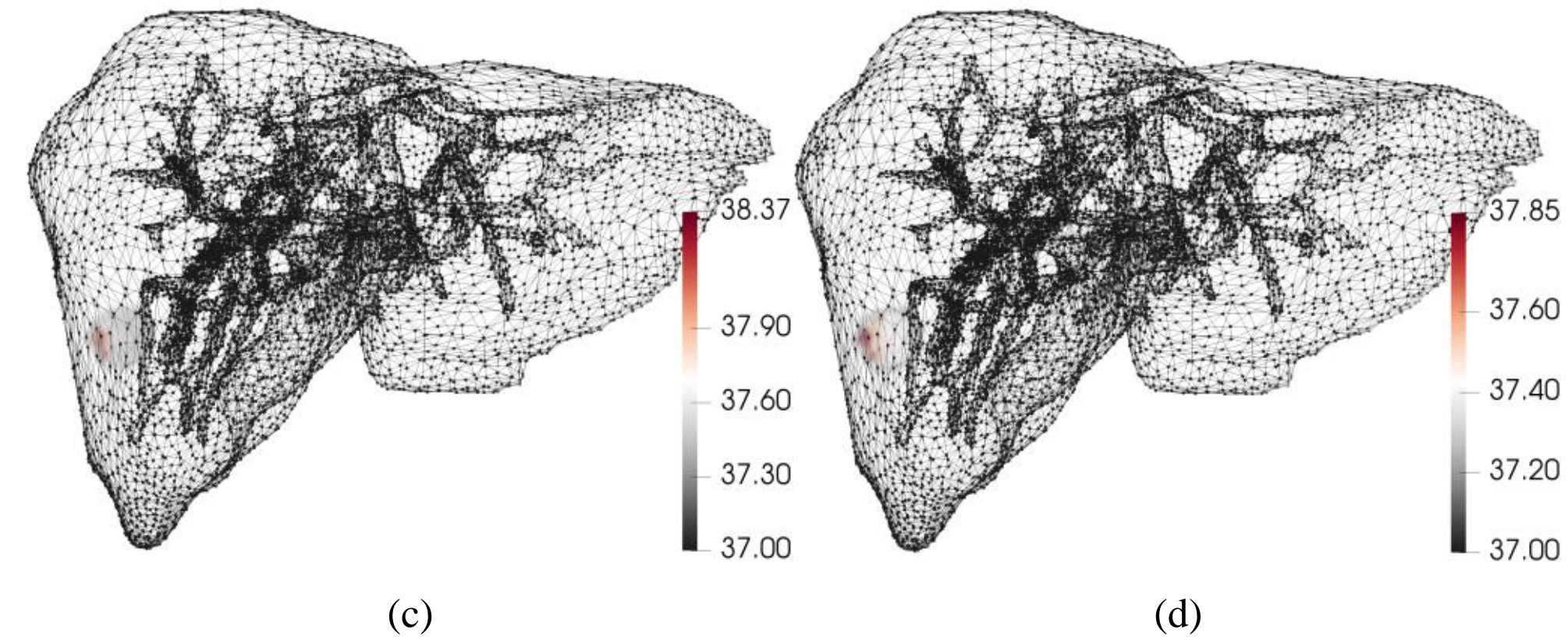

Fig. 3. Temperature distributions on the tested human liver model with embedded blood vessels by the proposed methodology at (a) $t = 5\ s$, (b) $t = 10\ s$, (c) $t = 15\ s$ and (d) $t = 20\ s$.

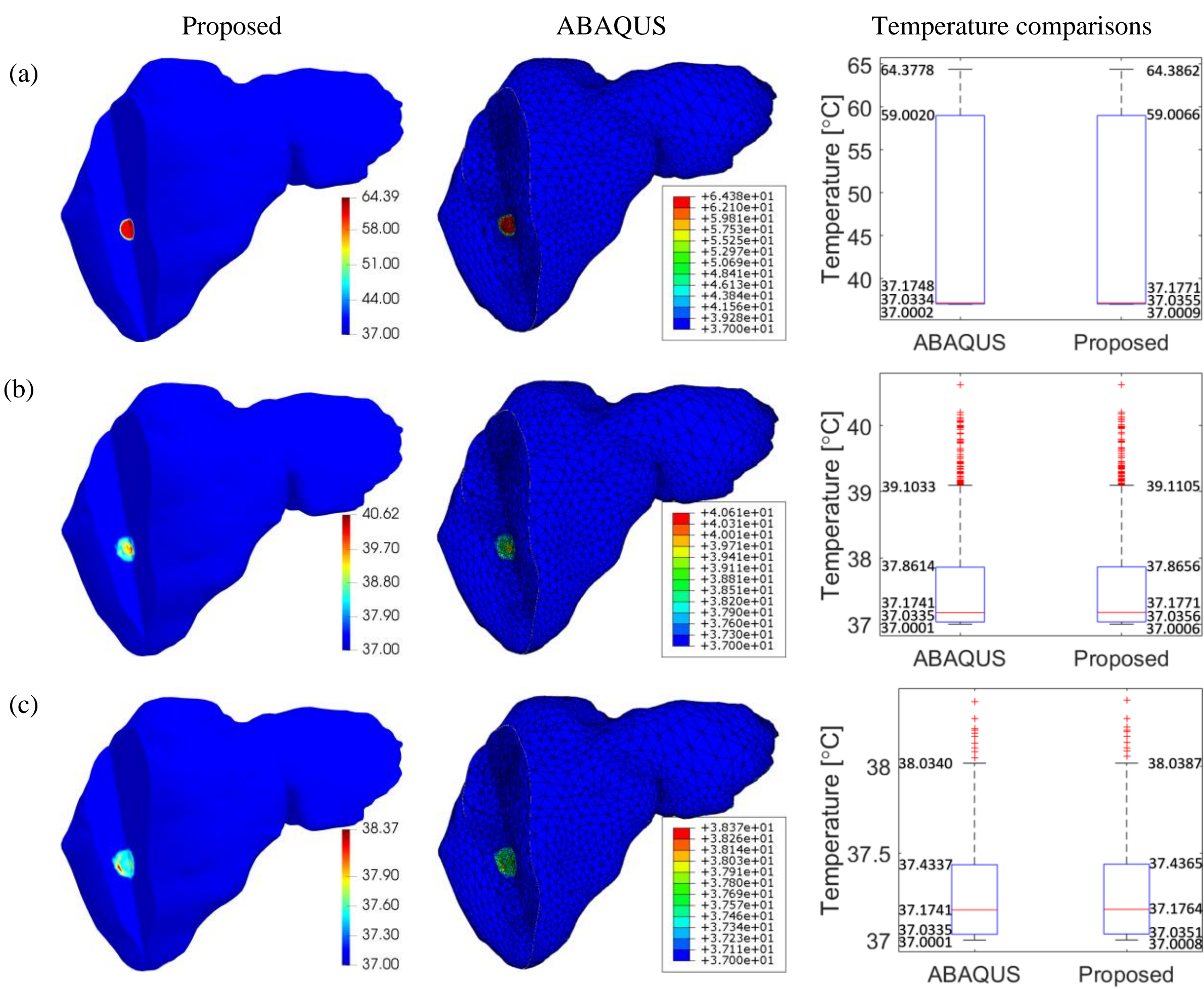

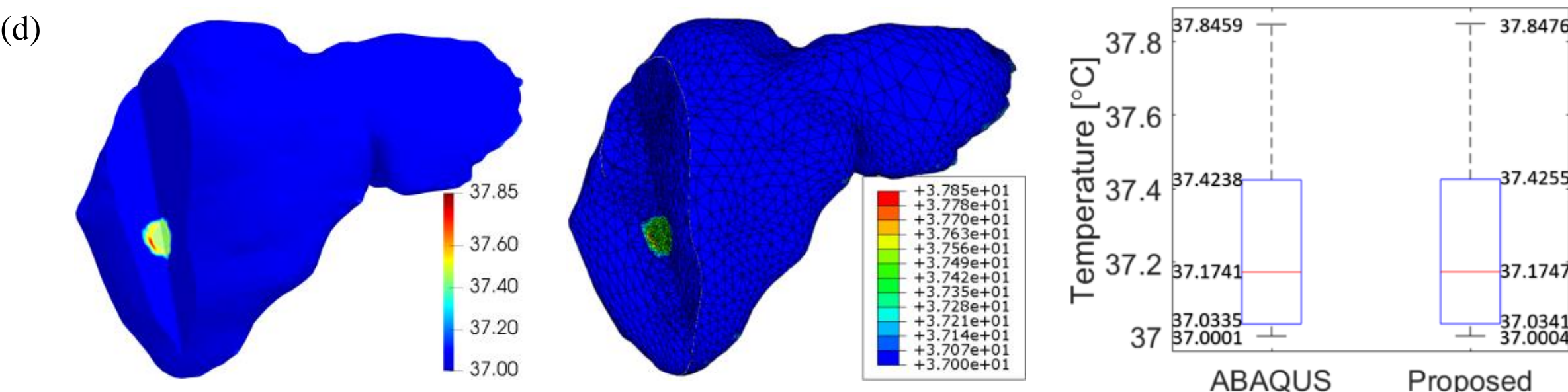

Fig. 4. Temperature distributions by the proposed methodology and ABAQUS and comparative analyses at nodes with temperature variations at (a) $t = 5\ s$, (b) $t = 10\ s$, (c) $t = 15\ s$ and (d) $t = 20\ s$.

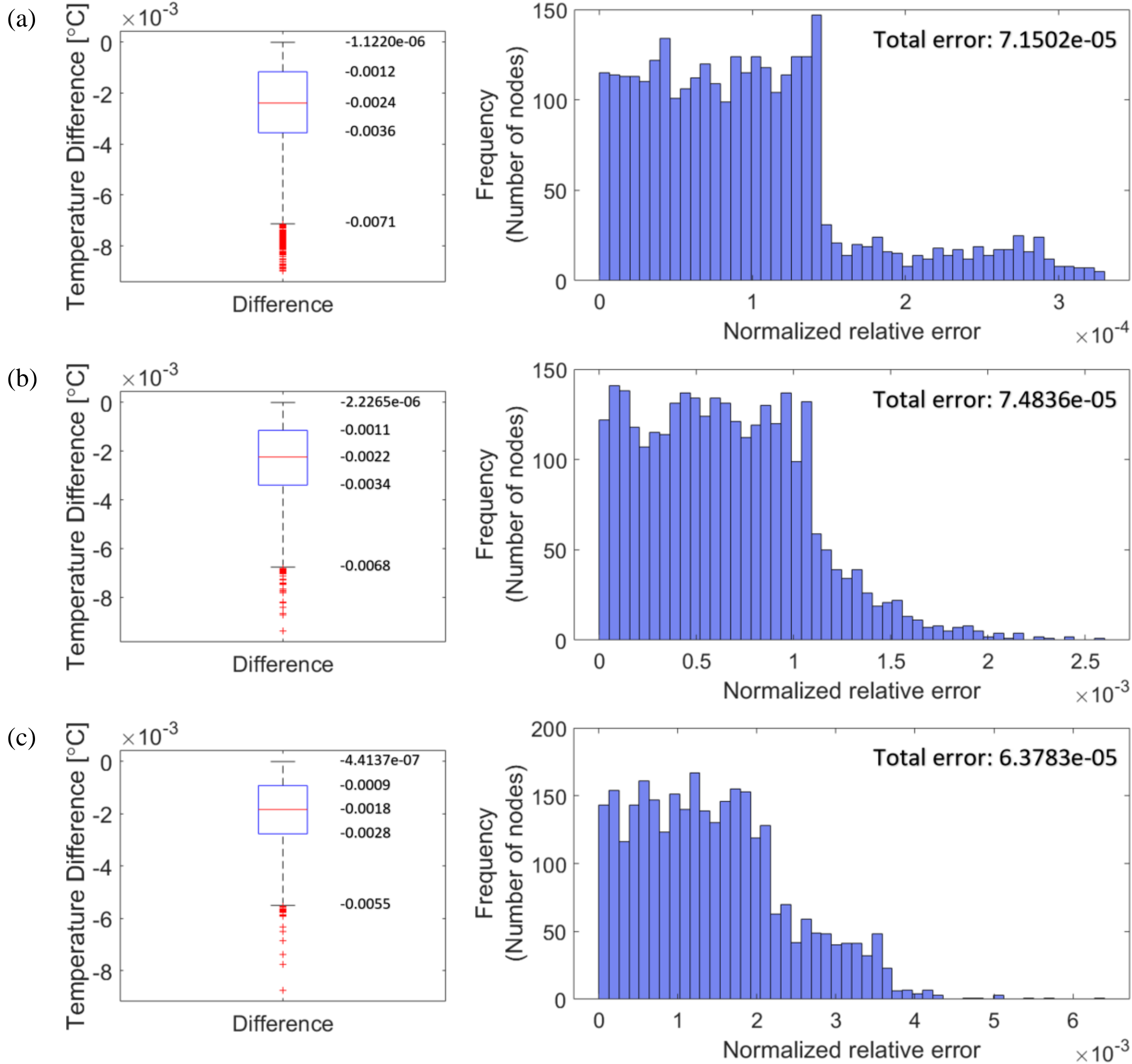

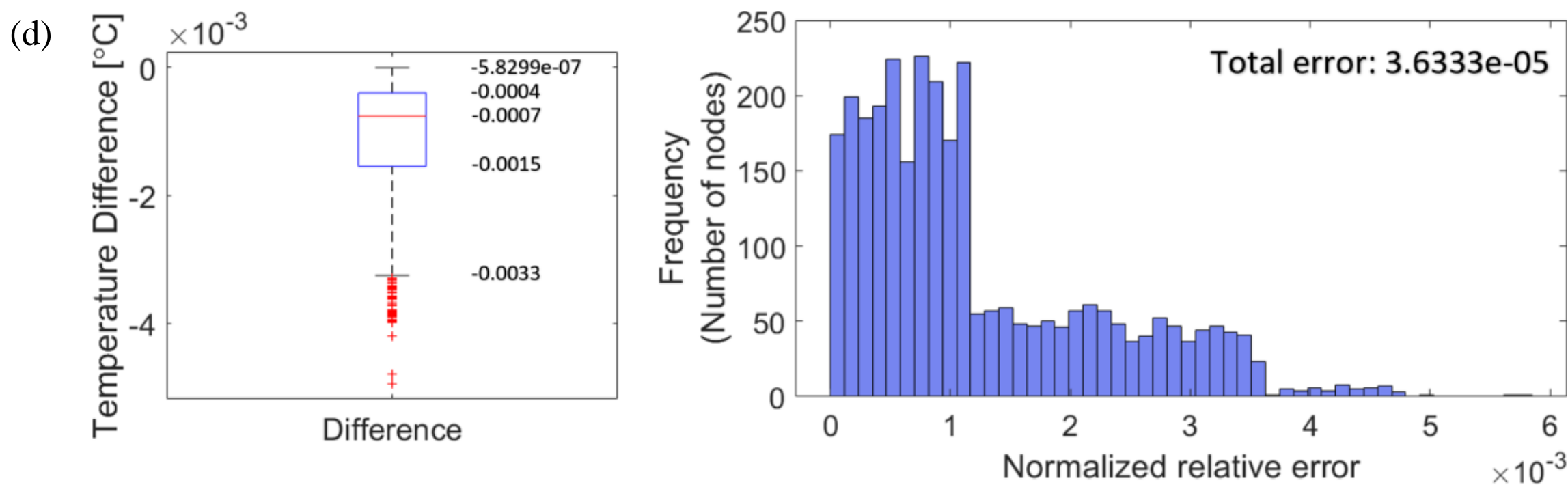

Fig. 5. Temperature differences ($T^{ABAQUS} - T^{Proposed}$) and the histogram displaying normalized relative errors node by node for the tested human liver model for temperature distributions at (a) $t = 5\ s$, (b) $t = 10\ s$, (c) $t = 15\ s$ and (d) $t = 20\ s$.

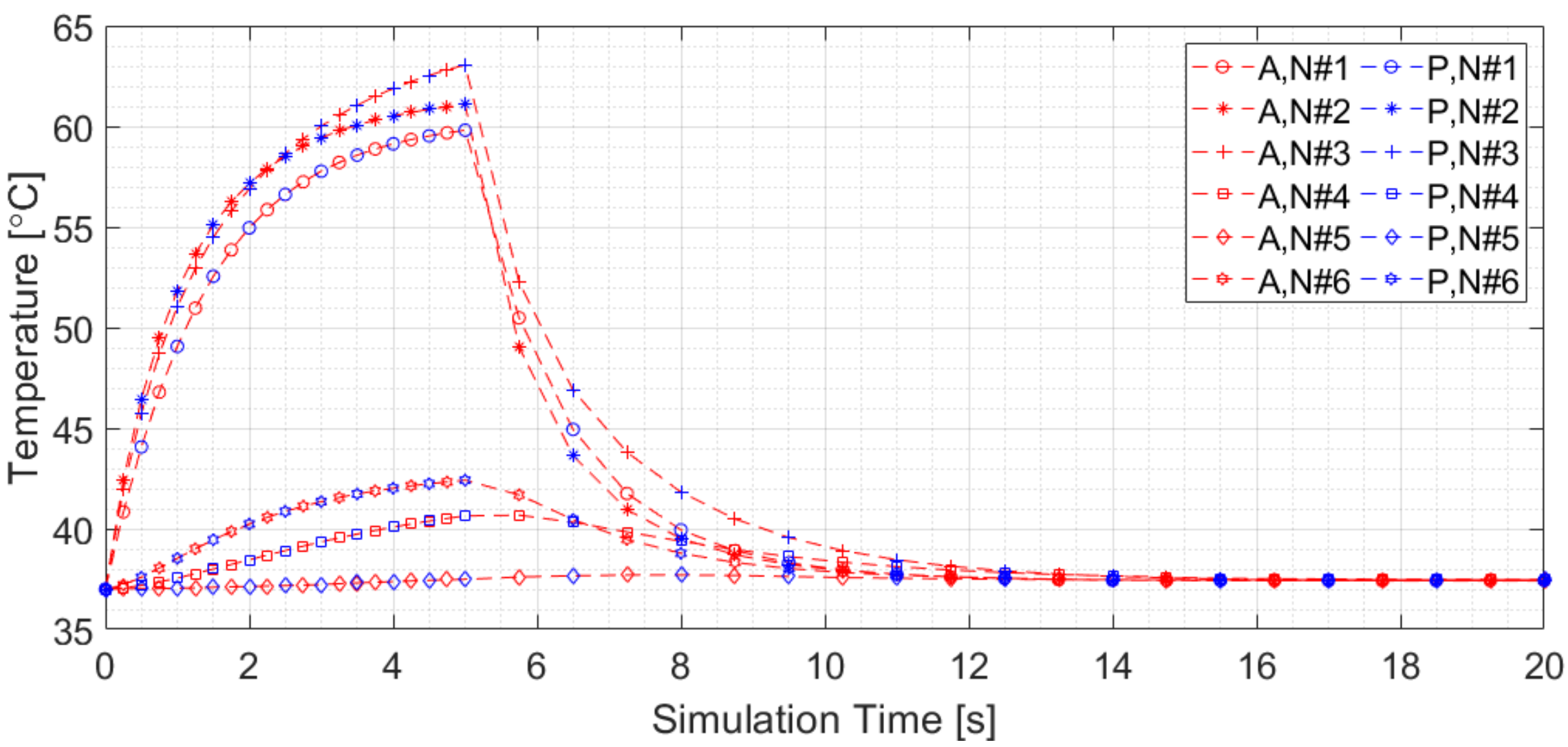

Fig. 6. Comparison of temperature-history between the results of the proposed method (P,N#) and ABAQUS (A,N#) at selected nodes.

## 4.2 Computational performance

The proposed PBHT based on FED-FEM considering soft tissue deformation is implemented in C++ and evaluated on an Intel(R) Core(TM) i7-8750H CPU @ 2.20 GHz and 16.0 GB RAM Laptop using a serial execution. A model discretized into six mesh densities, (i, 1445 nodes, 6738 elements), (ii, 1840 nodes, 8724 elements), (iii, 2644 nodes, 12852 elements), (iv, 3339 nodes, 16371 elements), (v, 5304 nodes, 26550 elements), and (vi, 7872 nodes, 40021 elements), are used to evaluate the computation time. The model has constant thermal and mechanical material properties. The simulation is performed at a physical time step size $\Delta t = 0.001\ s$ for 1000 time steps for simulation of $t = 1.0\ s$ transient non-linear PBHT process under soft tissue deformation. The deformable solver for mechanical deformation is a Saint Venant-Kirchhoff hyperelastic model [53]. Fig. 7 presents computation times of the model by the proposed methodology showing total computation time (Total), time of computing deformation (Mechanical), and time of computing proposed bio-heat transfer (Thermal-proposed). The proposed methodology is able to perform a calculation (Total) at $t = 14.110\ ms$ per time step with a standard error $t = \pm 0.09395\ ms$, which is composed of a calculation of soft tissue deformation (Mechanical) at $t = 6.296\ ms$ per time step with a standard error $t = \pm 0.06214\ ms$ and a calculation of bio-heat transfer (Thermal-proposed) at $t = 7.814\ ms$ per time step with a standard error $t = \pm 0.1089\ ms$, at (7872 nodes, 40021 elements) and completes the simulation (1000 time steps) in $14110.4\ ms$.

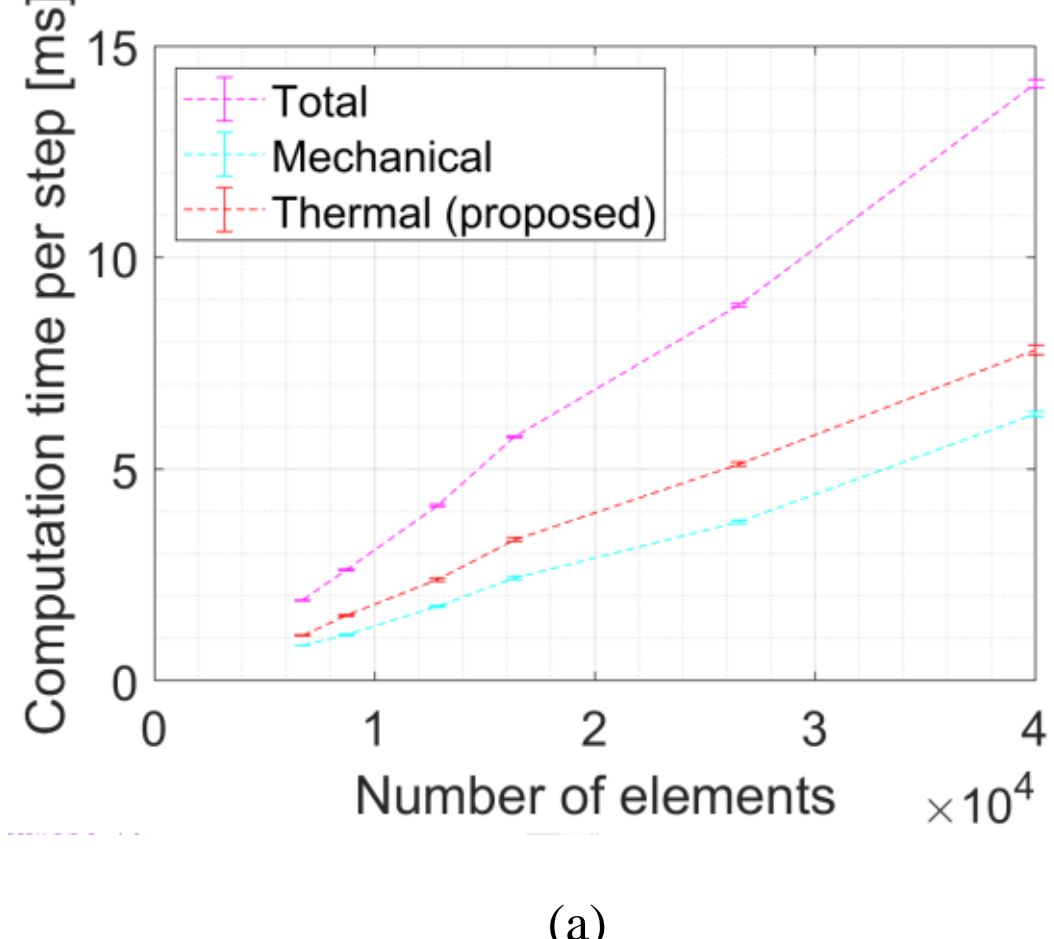

(a)

| Elements | Total [ms] | Mechanical [ms] | Thermal (proposed) [ms] |
|---|---|---|---|
| 6738 | 1902.1 | 834.5 | 1067.6 |
| 8724 | 2619.5 | 1083.6 | 1535.9 |
| 12852 | 4139.6 | 1755.8 | 2383.8 |
| 16371 | 5765.3 | 2430.2 | 3335.1 |
| 26550 | 8872.8 | 3753.6 | 5119.2 |
| 40021 | 14110.4 | 6296.7 | 7813.7 |
| 1000 time steps | | | |

(b)

Fig. 7. Computation times of different mesh sizes showing total computation time (Total), time of computing soft tissue deformation (Mechanical) and time of computing the proposed bio-heat transfer (Thermal-proposed): (a) computation time per time step and (b) total computation time (1000 time steps).

The computation is further conducted to investigate the computational effort of the new formulation for element nodal thermal loads computation. The following formulations of four-node linear tetrahedral element nodal thermal loads are investigated:

(i) Proposed PBHT under soft tissue deformation

$$ {}^t_0\hat{\mathbf{F}}_e = {}^0V_{tet}\left(\mathcal{L}\,{}^0\mathbf{h}_{\mathbf{x}}\,{}^t_0\mathbf{F}^{-1}\right)^T\,{}^t\mathbf{D}\left(\mathcal{L}\,{}^0\mathbf{h}_{\mathbf{x}}\,{}^t_0\mathbf{F}^{-1}\right)\det\left({}^t_0\mathbf{F}\right)\,{}^t\mathbf{T}(\mathbf{x}) \tag{20} $$

(ii) Classical PBHT

$$ {}^t_0\hat{\mathbf{F}}_e = \underbrace{{}^0V_{tet}\left(\mathcal{L}\,{}^0\mathbf{h}_{\mathbf{x}}\right)^T}_{pre-computed}\,{}^t\mathbf{D}\left(\mathcal{L}\,{}^0\mathbf{h}_{\mathbf{x}}\right)\,{}^t\mathbf{T}(\mathbf{x}) \tag{21} $$

(iii) Classical PBHT with temperature-independent thermal conductivity ${}^0\mathbf{D}$

$$ {}^t_0\hat{\mathbf{F}}_e = \underbrace{{}^0V_{tet}\left(\mathcal{L}\,{}^0\mathbf{h}_{\mathbf{x}}\right)^T\,{}^0\mathbf{D}\left(\mathcal{L}\,{}^0\mathbf{h}_{\mathbf{x}}\right)}_{pre-computed}\,{}^t\mathbf{T}(\mathbf{x}) \tag{22} $$

(iv) Classical PBHT with isotropic thermal conductivity ${}^tk$

$$ {}^t_0\hat{\mathbf{F}}_e = {}^tk\,\underbrace{{}^0V_{tet}\left(\mathcal{L}\,{}^0\mathbf{h}_{\mathbf{x}}\right)^T\left(\mathcal{L}\,{}^0\mathbf{h}_{\mathbf{x}}\right)}_{pre-computed}\,{}^t\mathbf{T}(\mathbf{x}) \tag{23} $$

(v) Classical PBHT with isotropic temperature-independent thermal conductivity ${}^0k$

$$ {}^t_0\hat{\mathbf{F}}_e = \underbrace{{}^0k\,{}^0V_{tet}\left(\mathcal{L}\,{}^0\mathbf{h}_{\mathbf{x}}\right)^T\left(\mathcal{L}\,{}^0\mathbf{h}_{\mathbf{x}}\right)}_{pre-computed}\,{}^t\mathbf{T}(\mathbf{x}) \tag{24} $$

Fig. 8 illustrates a comparison of computation times of the above formulations for one element nodal thermal loads computation. Formulation (i) considering soft tissue deformation consumes $t = 2.518 \times 10^{-4}\ ms$ with a standard error $t = \pm 4.52 \times 10^{-6}\ ms$. Formulation (ii) consumes $t = 2.237 \times 10^{-4}\ ms$ with a standard error $t = \pm 2.40 \times 10^{-6}\ ms$, which is 0.89 times of formulation (i) due to pre-computation of the first two terms. Formulation (iii) consumes 0.57 times of formulation (i) and is matched by formulation (v) (0.58 times) since both formulations can pre-compute the full thermal stiffness. Formulation (iv) consumes 0.74 times of formulation (i) and consumes more time than formulations (iii) and (v) since the thermal conductivity needs to be updated at each time step.

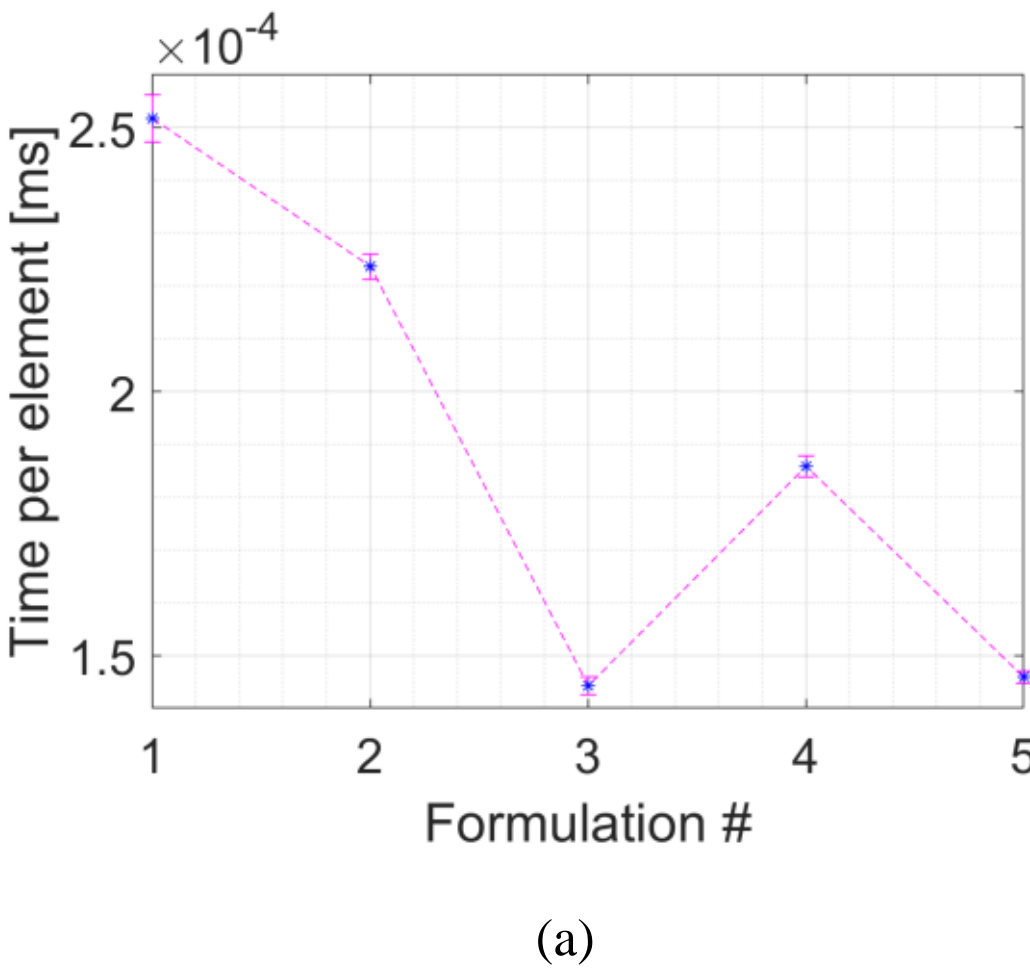

(a)

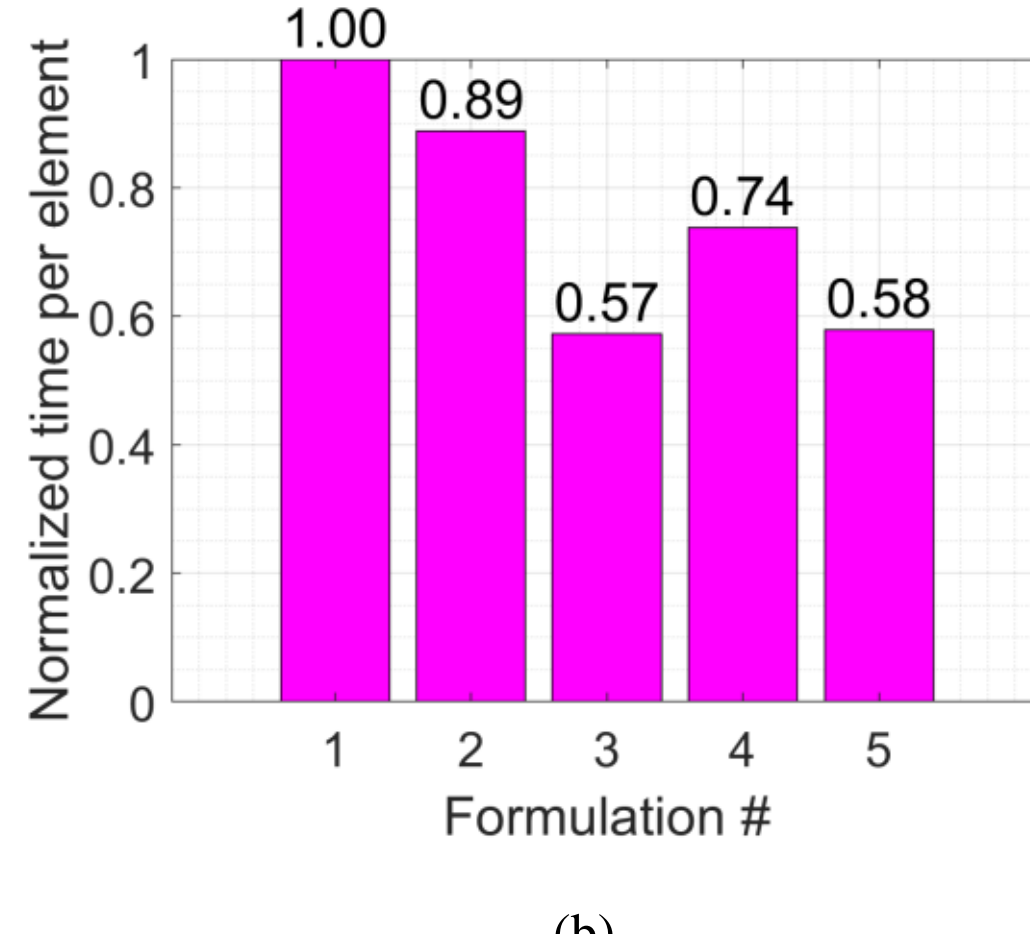

(b)

Fig. 8. Comparison of computation times of the five formulations (Eqs. (20-24)) of the four-node linear tetrahedral element nodal thermal loads: (a) computation times per element and (b) normalized computation times per element.

## 5. Discussion

The classical PBHT exists certain simplifications and assumptions. The blood flow in the capillaries is assumed isotropic and hence the directional-dependent blood flow heat transfer is not modelled [19, 54]. The physical processes such as water evaporation and transport of vapor are not captured in the classical PBHT [25, 37, 55]; however, the phase change occurs when tissue temperature elevates beyond the vaporisation threshold $T_{thres}$ ($T_{thres} = 100℃$ [36]), but the maximum temperature reached in the present work is less than $T = 65℃$ (see Fig. 4(a)). The present work enforces Dirichlet boundary conditions to the blood flowing region based on the findings from studies [14, 49] for reasonable vessel wall temperature, and a more realistic prediction may be obtained by incorporating blood flow velocity [5, 48] but at the cost of additional computation complexity. In general, despite the development of a more complex and rigorous bio-heat transfer model, PBHT still remains an effective model for bio-heat transfer analysis during thermal heating procedures [25], whereas other models can be used but at the expense of added numerical complexity and computational load.

The proposed method employs explicit scheme for time integration in the temporal domain, and hence precautions must be taken for time increment size to meet the Courant-Friedrichs-Lewy (CFL) condition [56] for numerical stability. The critical time step is determined by $\Delta t_{cr} = \frac{2}{|\lambda_{max}|}$ where $\lambda_{max}$ is the largest eigenvalue of the global thermal stiffness over thermal mass values [30], but its computation can be very expensive when thermal matrices are large. Rong et al. [57] provides a method for estimation of the largest eigenvalues and critical time steps, and it was estimated $\Delta t_{cr} \approx 4.435 \times 10^{-5}\ ms$ for the proposed method at the undeformed state. However, a smaller critical time step $\Delta t_{cr} \approx 4 \times 10^{-5}\ ms$ was used throughout the simulation for numerical stability at the deformed state of soft tissue. Due to soft tissue deformation the geometry of finite elements will be changed during simulation which can lead to variation in critical time step sizes [58] for proposed PBHT simulation, and they can be difficult to estimate when complex non-linear analyses are considered [59]. Furthermore, the discretization of simulated models will affect the critical time step, in particular, the blood vessel networks in the liver which require a fine discretization for the mesh. The critical time step decreases as the mesh becomes finer for the blood vessels, increasing the number of simulation steps that prolongs the total computation time; therefore, a careful selection of the included blood vessels and other fine details of the liver needs to be considered for achieving a balance of critical time steps for efficient simulation. For numerical stability, it was reported that neural networks [60, 61], adaptive semi-explicit/explicit time marching [59] and unconditionally stable explicit scheme [62] could achieve stable numerical analysis for non-linear dynamic systems; these methods may be incorporated into the proposed method for stable dynamic bio-heat transfer analysis under soft tissue deformation.

The proposed methodology can be employed as a standalone thermal solver in additional to existing real-time soft tissue deformable algorithms [45] with nodal displacements and time step sizes in sync for computation of soft tissue thermal response under tissue deformation. It conducts thermal analysis under deformation by incorporating nodal displacements for deformation gradient from the mechanical computation, and the thermal and mechanical solvers are performed individually, allowing various mechanical material properties (e.g., linear elasticity, hyperelasticity, anisotropy, viscosity and compressibility [52]) to be simulated at small or large deformation of soft tissue. The proposed method not only extends the current method [30] to incorporate soft tissue deformation but also constitutes a step towards dynamic (due to deformation) temperature optimization and thermal dosimetry computation, allowing interactive thermo-mechanical surgical training, fast temperature and thermal dose computation for treatment planning, and image tracking-based tissue temperature prediction and guidance for robot-assisted surgical interventions.

## 6. Conclusions

This paper presents a fast numerical formulation for efficient computation of thermal response of soft tissue subject to deformation based on FED-FEM algorithm. The proposed solution procedure extends the formulation of static non-moving classical PBHT to dynamic PBHT under soft tissue deformation via a transformation of spatial domain from the unknown deformed state to the known initial static non-moving state by a mapping function in the form of deformation gradient. Numerical accuracy and computational costs are evaluated on a realistic virtual human liver model with blood vessel networks embedded for a clinically relevant application of thermal ablative treatment of hepatic cancer. The proposed method can be employed in addition to existing real-time mechanical deformable models [45] with synchronized nodal displacements and time step sizes, and it can enable dynamic (due to deformation) tissue temperature analysis for surgical training, planning and guidance. Future research work will focus on the development of a dynamic temperature prediction system for image-guided robot-assisted thermal therapy [63]. The deformation of soft tissue may be captured by medical imaging devices, and the tissue temperature field can be obtained on the deformed tissue to facilitate surgical feedback during the thermo-therapeutic treatment.

## Acknowledgment

This work is funded by the National Health and Medical Research Council (NHMRC), Australia, Grant 1093314.